\documentclass[english,twocolumn,superscriptaddress, aps, prc]{revtex4-2}
\usepackage[T1]{fontenc}
\usepackage[utf8]{inputenc}
\usepackage{color}
\usepackage{babel}
\usepackage{amsmath}
\usepackage{graphicx}
\usepackage[normalem]{ulem}

\usepackage[unicode=true,pdfusetitle,
 bookmarks=true,bookmarksnumbered=false,bookmarksopen=false,
 breaklinks=false,pdfborder={0 0 1},backref=false,colorlinks=true]
 {hyperref}
\hypersetup{
 linkcolor=blue, citecolor=cyan}

\makeatletter
\usepackage{amsthm}

\usepackage{xcolor}

\makeatother

\begin{document}
\title{Global polarization of $\Lambda$ hyperons and its sensitivity \\
to equations of state  in low-energy heavy-ion collisions}

\author{Cong Yi}
\email{congyi@mail.ustc.edu.cn}

\affiliation{Institute of Particle Physics and Key Laboratory of Quark and Lepton
Physics (MOE), Central China Normal University, Wuhan 430079, China}

\author{Shi Pu}
\email{shipu@ustc.edu.cn}
\affiliation{Department of Modern Physics and Anhui Center for fundamental Sciences
(Theoretical Physics), University of Science and Technology of China,
Anhui 230026}
\affiliation{Southern Center for Nuclear-Science Theory (SCNT), Institute of Modern
Physics, Chinese Academy of Sciences, Huizhou 516000, Guangdong Province,
China}

\author{Long-Gang Pang}
\email{lgpang@ccnu.edu.cn}
\affiliation{Institute of Particle Physics and Key Laboratory of Quark and Lepton
Physics (MOE), Central China Normal University, Wuhan 430079, China}

\author{Guang-You Qin}
\email{guangyou.qin@ccnu.edu.cn}
\affiliation{Institute of Particle Physics and Key Laboratory of Quark and Lepton
Physics (MOE), Central China Normal University, Wuhan 430079, China}

\author{Xin-Nian Wang}
\email{xnwang@lbl.gov}
\affiliation{Institute of Particle Physics and Key Laboratory of Quark and Lepton
Physics (MOE), Central China Normal University, Wuhan 430079, China}

\begin{abstract}
Significant global polarization of $\Lambda$ hyperons along the direction of the orbital angular momentum has been measured in non-central heavy-ion collisions where the equation of state (EOS) of the produced dense matter is expected to change from intermediate to low colliding energies. We study the sensitivity of the global $\Lambda$ polarization to EOS in heavy-ion collisions within the SMASH transport model. Among the three different EOS we considered, only the hadron resonance gas (HRG) describes the experimental data well at low colliding energies even when it is below the $\Lambda$ production threshold in nucleon-nucleon collisions. The polarization induced by thermal vorticity as a function of centrality, rapidity, and transverse momentum at $\sqrt{s_{NN}} = 3$ GeV in Au+Au collisions is shown to agree well with the experimental data. Our study also indicates a possible peak in the global $\Lambda$ polarization around $\sqrt{s_{NN}} = 2.4$ GeV in Au+Au collisions. Furthermore, we find that the rapidity and transverse momentum-dependent helicity polarization induced by thermal vorticity vanishes due to space-reversal symmetry.
\end{abstract}
\maketitle

\section{Introduction}

 A significant breakthrough in high-energy nuclear physics has been the observation of the global polarization of $\Lambda$ and $\overline{\Lambda}$ hyperons \citep{STAR:2017ckg,STAR:2007ccu,STAR:2018gyt,ALICE:2019onw, STAR:2021beb,HADES:2022enx} in non-central relativistic heavy-ion collisions. In these collisions, two nuclei are accelerated to nearly the speed of light and subsequently collide with large initial orbital angular momentum. During the collision process, part of initial orbital angular momentum is transferred to quarks via spin-orbit coupling, ultimately leading to the global polarization of hyperons \citep{Liang:2004ph,Gao:2007bc}. 

In spin fluid dynamics, one can link the global polarization of hyperons to the vorticity of the quark-gluon plasma (QGP) produced in these collisions. Using statistical quantum field theory \citep{Becattini:2013fla} and quantum kinetic theory \citep{Fang:2016vpj}, the global polarization of $\Lambda$ and $\overline{\Lambda}$ hyperons can be expressed as,
$\boldsymbol{P}_\Lambda + \boldsymbol{P}_{\overline{\Lambda}} \approx\boldsymbol{\omega}/T$, where $\boldsymbol{\omega}$ represents the averaged kinetic vorticity of the fluid, and $T$ denotes the effective temperature. This relation allows the vorticity of the QGP to be estimated from experimental data, revealing that the QGP is the \emph{most vortical fluid} ever observed \citep{STAR:2017ckg}. 

Significant progress has been made in quantitatively understanding the observed polarization of $\Lambda$ hyperons in high energy heavy-ion collisions. Numerous studies 
\citep{Becattini:2007nd,Betz:2007kg,Becattini:2013fla,Becattini:2013vja,Csernai:2013bqa,Fang:2016vpj,Karpenko:2016jyx,Xie:2017upb,Li:2017slc,Sun:2017xhx,Shi:2017wpk,Xia:2018tes,Shi:2019wzi,Fu:2020oxj,Lei:2021mvp,Ambrus:2020oiw,Vitiuk:2019rfv} (see also Refs.~\citep{Gao:2020vbh,Hidaka:2022dmn,Becattini:2024uha} and references therein) have found good agreement between model calculations and experimental data at medium and high collision energies. However, the behavior of the polarization at low-energy collisions remains not well understood. Two questions still remain unanswered: \emph{ Is thermal vorticity of the fluid still the underlying mechanism for the global polarization of $\Lambda$ hyperons in low-energy collisions? What is the underlying effective degrees of freedom and the EoS?}

If the global polarization in low  energy  collisions is still driven by thermal vorticity or other collective motion of the produced dense matter, one would expect the  polarization of hyperons to be approximately vanishing when the collision energy drops below $2m_N$, where $m_N$ is the nucleon mass. In this regime, the initial orbital angular momentum of the produced dense matter is expected to be strongly suppressed \citep{Deng:2020ygd, Deng:2021miw}, thus the medium cannot generate substantial vorticity.
On the other hand, current data from STAR \cite{STAR:2021beb} and HADES \cite{HADES:2022enx} show an increasing global polarization as the collision energy decreases. In Ref.~\cite{Guo:2021udq}, it was found that global polarization calculated using models based on kinetic theory fails to describe the experimental data. Subsequent studies \cite{Deng:2020ygd,Deng:2021miw,Vitiuk:2019rfv} found that results from other transport models, while subject to large uncertainties, can reproduce the data. 
These results challenge the conjecture that thermal vorticity alone induces global $\Lambda$ hyperon polarization in low-energy collisions. In addition, recent works \citep{Sun:2025oib,Liu:2025kpp,Zheng:2025ngn,Xu:2026hxz} have extended the studies to spin polarization of nucleons and  hypernuclei.
Interestingly, a recent study \cite{Ivanov:2020udj} employing extended hydrodynamic models suggests that the global polarization in low-energy collisions may depend on the equation of state (EOS). 
However, whether such dependence persists within the framework of transport models remains unexplored.
It is therefore necessary to revisit the  global polarization in low-energy collisions using transport models and to systematically investigate its sensitivity to the effective EOS. 


In this work, we employ the transport model, Simulating Many Accelerated Strongly-interacting Hadrons (SMASH)~\cite{Weil:2016zrk}, with the modified Cooper-Frye formula~\citep{Becattini:2013fla, Fang:2016vpj} to investigate global polarization of $\Lambda$ hyperons in low-energy collisions. The remainder of this paper is organized as follows. We begin by introducing our theoretical model and numerical setup. Next, we present our numerical results and analyze their dependence on the equation of state, centrality, rapidity, and transverse momentum. Finally, we discuss the helicity polarization, in low-energy collisions and provide the conclusions.

\section{Theoretical framework}

The modified Cooper-Frye formula \citep{Becattini:2013fla, Fang:2016vpj} is the commonly used theoretical framework for computing the spin polarization of spin-$1/2$ particles. In this framework, the polarization pseudovector in the laboratory frame is expressed as:
\begin{eqnarray}
\mathcal{S}^\mu(x,p) = \frac{1}{8m_\Lambda}(1-f)\varepsilon^{\mu\nu\alpha\beta}p_\nu \varpi_{\alpha\beta}, \label{eq:Smu}
\end{eqnarray}
where
\begin{eqnarray}
\varpi_{\alpha\beta} = \frac{1}{2} \left[\partial_{\alpha}\left(\frac{u_{\beta}}{T}\right) - \partial_{\beta}\left(\frac{u_{\alpha}}{T}\right)\right], \label{eq:th vorticity}
\end{eqnarray}
is the thermal vorticity, $f$ is the particle distribution function, $p^\mu=(E_p,\boldsymbol{p})$ and $m_\Lambda$ is the mass of the $\Lambda$ hyperon. For simplicity, $1-f \simeq 1$ is often assumed in the above formula.

To derive Eq.~\eqref{eq:Smu}, it is assumed that the system has reached global equilibrium. When the system is instead at local equilibrium, various hydrodynamic effects, such as shear-induced polarization \citep{Hidaka:2017auj,Liu:2020dxg,Becattini:2021suc,Liu:2021uhn,Fu:2021pok,Becattini:2021iol,Yi:2021ryh,Ryu:2021lnx,Florkowski:2021xvy,Buzzegoli:2022fxu,Becattini:2022zvf,Palermo:2022lvh,Wu:2022mkr,Fu:2022myl,Fu:2022oup,Palermo:2024tza}, the baryonic spin Hall effect \citep{Hidaka:2017auj,Yi:2021ryh,Liu:2020dxg,Liu:2021uhn,Becattini:2021suc,Fu:2021pok,Fu:2022myl,Becattini:2021iol,Wu:2022mkr}, and other novel spin transport  \cite{Fang:2022ttm,Fang:2023bbw, Fang:2024vds,Fang:2025pzy, Wang:2025mfz,Wang:2022yli,Lin:2022tma,Yamamoto:2023okm,Lin:2024zik}, can contribute to the polarization pseudovector. However, it has been shown that, all these hydrodynamic effects, except for the thermal vorticity, approximately vanish in their contribution to the global polarization in medium- and high-energy collisions.

We emphasize that local equilibrium may not be achieved in very low-energy collisions. If the system deviates from local equilibrium, additional contributions from non-equilibrium effects may arise. However, based on previous studies in the high-energy limit, certain non-equilibrium effects do not contribute to global polarization. For instance, corrections from second-order terms in the gradient expansion ~\cite{Fang:2024vds} were found to have no impact on global polarization. Therefore, we neglect further corrections arising from deviations beyond local equilibrium in this studies. This assumption will be revisited in the conclusion.

To compare with experimental data, the spin polarization vector needs to be transformed into the local rest frame of the $\Lambda$ hyperons as,
\begin{eqnarray}
\mathbf{S}^{*}(x,p) = \mathbf{S} - \frac{\mathbf{p} \cdot \mathbf{S}}{E_{p}(m_\Lambda+E_{p})} \mathbf{p},
\end{eqnarray}
where $\mathbf{S}$ is the spin polarization vector in the laboratory frame as shown in Eq.~\eqref{eq:Smu}.
The averaged spin polarization vector is then defined as:
\begin{eqnarray}
\langle \mathbf{P}^{i} \rangle = \frac{1}{N} \sum_{i=1}^{N} \frac{\mathbf{S}_{i}^{*}(x_{i},p_{i})}{S},
\end{eqnarray}
where $N$ is the total number of $\Lambda$ hyperons, and $i = 1, 2, 3$ corresponds to the spin polarization along the $x$, $y$, and $z$ directions, respectively. Here, $y$ denotes the direction of the system’s angular momentum, $z$ is the beam direction, and $x$ is perpendicular to the $y$–$z$ plane. Our polarization pseudovector is particle-number-weighted, different with the energy-density-weighted approach used in Refs.~\cite{Deng:2020ygd, Deng:2021miw}.

\section{Numerical setups}
We employ the SMASH transport model~\cite{Weil:2016zrk} to simulate the evolution of the collision system and calculate the thermal vorticity. SMASH is a Boltzmann-Uehling-Uhlenbeck (BUU)-type transport model that solves the relativistic hadronic Boltzmann equation:
\begin{eqnarray}
p^{\mu}\partial_{\mu}f_{i}(x,p) + m_{i}F^{\alpha}\partial_{\alpha}^{p}f_{i}(x,p) = C_{\text{coll}}^{i},
\label{eq:SMASH_BE}
\end{eqnarray}
where the index $i$ represents different hadron species. The model incorporates all hadrons listed by the Particle Data Group~\cite{ParticleDataGroup:2020ssz} with masses up to $m_{i} \sim 2.35$ GeV. Here, $f_{i}$ is the phase-space distribution function for hadron species $i$, and $C_{\text{coll}}^{i}$ denotes the collision term, which accounts for string fragmentation, elastic and inelastic collisions, as well as resonance decay and formation. The term $F^{\mu}$ represents the external mean-field force.

In low-energy collisions, in addition to particle collisions, mean-field potentials also influence particle motion.  The spatial gradients of the mean-filed potentials is related to the effective force $F^\alpha$ in Eq.~(\ref{eq:SMASH_BE}). Therefore, in our simulations, finite mean-field potentials are included when the collision energy is below $\sqrt{s_{NN}} \leq 3$ GeV. The potentials are parameterized as follows~\cite{Weil:2016zrk}:
\begin{eqnarray}
U_{sk} & = & A\left(\frac{\rho}{\rho_{0}}\right)+B\left(\frac{\rho}{\rho_{0}}\right)^{\tau}, \nonumber \\
U_{Sym} & = & \pm S_{pot}\frac{I_{3}}{I}\frac{\rho_{I_{3}}}{\rho_{0}},
\end{eqnarray}
where $U_{sk}$ and $U_{Sym}$ are the isoscalar Skyrme term and the isovector symmetry term of the mean-field potential, respectively. Here, $\rho$ and $\rho_{I3}$ represent the net baryon density and baryon isospin density associated with the relative isospin projection $I_{3}/I$, respectively, defined in the Eckart rest frame. The nuclear ground-state density is set to $\rho_{0} = 0.168~\text{fm}^{-3}$.
Following the default parameters in SMASH, we use $A = -209.2$ MeV, $B = 156.4$ MeV, and $\tau = 1.35$ to characterize the attraction, repulsion, and stiffness of the nuclear matter. The symmetry potential parameter is set to $S_{pot} = 18$ MeV. This parameter set successfully reproduces the soft particle yields at $\sqrt{s_{NN}} = 3$ GeV~\cite{Lin:2025dsm}. We also checked numerically that the contribution of mean-field effects to the global polarization is approximately negligible when the collision energy $\sqrt{s_{NN}} > 3$ GeV.

In current work, the impact parameter
$b$ is mapped to centrality $c$ using the geometric approximation
$b\approx2R_{Au}\sqrt{c}$ with $R_{Au}$ being the radius of Au atom. The $20$–$50\%$ centrality corresponds  approximately to impact parameters $b \in [6, 9]$ fm in the transport model~\citep{Guo:2021udq,Deng:2020ygd,Deng:2021miw}. We compute the thermal vorticity by taking the event-by-event average over at least $10^5$ events for each collision energy, with the impact parameter randomly chosen within $[6, 9]$ fm. 
The thermal vorticity contains information about the fluid velocity and temperature at the points where the $\Lambda$ hyperons are generated. The details for extracting the fluid velocity and temperature are as follows.

\subsection{Fluid velocity, baryon number density and energy density}
After solving the relativistic hadronic Boltzmann equation~\eqref{eq:SMASH_BE}, the SMASH model provides the phase-space coordinates (position and momentum) of all particles. A coarse-graining method is then applied to extract the thermal vorticity field at different collision energies. The collision volume is divided into spatial cells with grid size $\Delta t = \Delta x = \Delta y = \Delta z = 1 \text{fm}$. The energy-momentum tensor and baryon current in each cell are calculated as~\cite{Li:2017slc}:
\begin{eqnarray}
T^{\mu\nu}(t,x,y,z) & = & \frac{1}{N_{e}\Delta V}\sum_{m=1}^{N_{e}}\sum_{n}\frac{p_{mn}^{\mu}p_{mn}^{\nu}}{E_{mn}}, \nonumber \\
J_{B}^{\mu}(t,x,y,z) & = & \frac{1}{N_{e}\Delta V}\sum_{m=1}^{N_{e}}\sum_{n}N_{B}^{mn}\frac{p_{mn}^{\mu}}{E_{mn}},
\end{eqnarray}
where $E_{mn}$, $p_{mn}^{\mu}$, and $N_{B}^{mn}$ denote the energy, four-momentum, and baryon number of the $n$-th particle in the $m$-th event, respectively. $\Delta V$ is the volume of each cell, and $N_{e}$ is the total number of events, which is at least $10^{5}$ for each energy point. 

The local flow velocity $u^{\mu}$ and energy density $\epsilon$ are defined as the eigenvector and eigenvalue of the energy-momentum tensor in the Landau frame,
\begin{eqnarray}
T^{\mu}_{\; \nu}u^{\nu} & = & \epsilon u^{\mu}, \nonumber \\
n_{B} & = & J_{B}^{\mu}u_{\mu},
\end{eqnarray}
where $n_{B}$ represents the local net baryon number density in each cell. 

\begin{figure}
\includegraphics[scale=0.43]{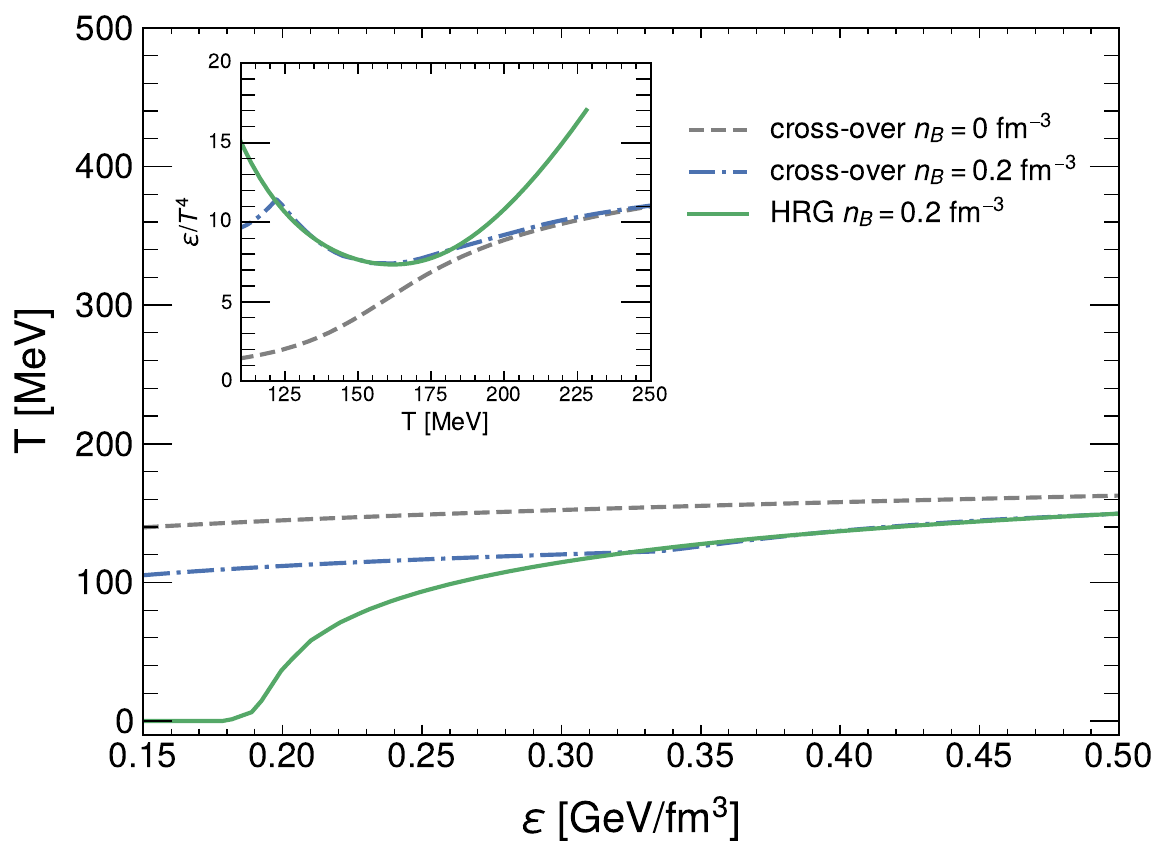}
\caption{An example of extracting the temperature $T$ as a function of energy density $\epsilon$ at a specific net baryon density  $n_{B}=0\; \text{fm}^{-3}$ or  $n_{B}=0.2\; \text{fm}^{-3}$. The gray dashed, blue dash-dotted, and green solid lines represent the extracted temperature using the HotQCD EOS, NEOS-BQS EOS 
and HRG EOS, respectively. The top-left subplot shows $\epsilon/T^4$ as a function of $T$ for the three EOS.
}
\label{fig:EOS}
\end{figure}


\begin{figure*}[t]
\includegraphics[scale=0.5]{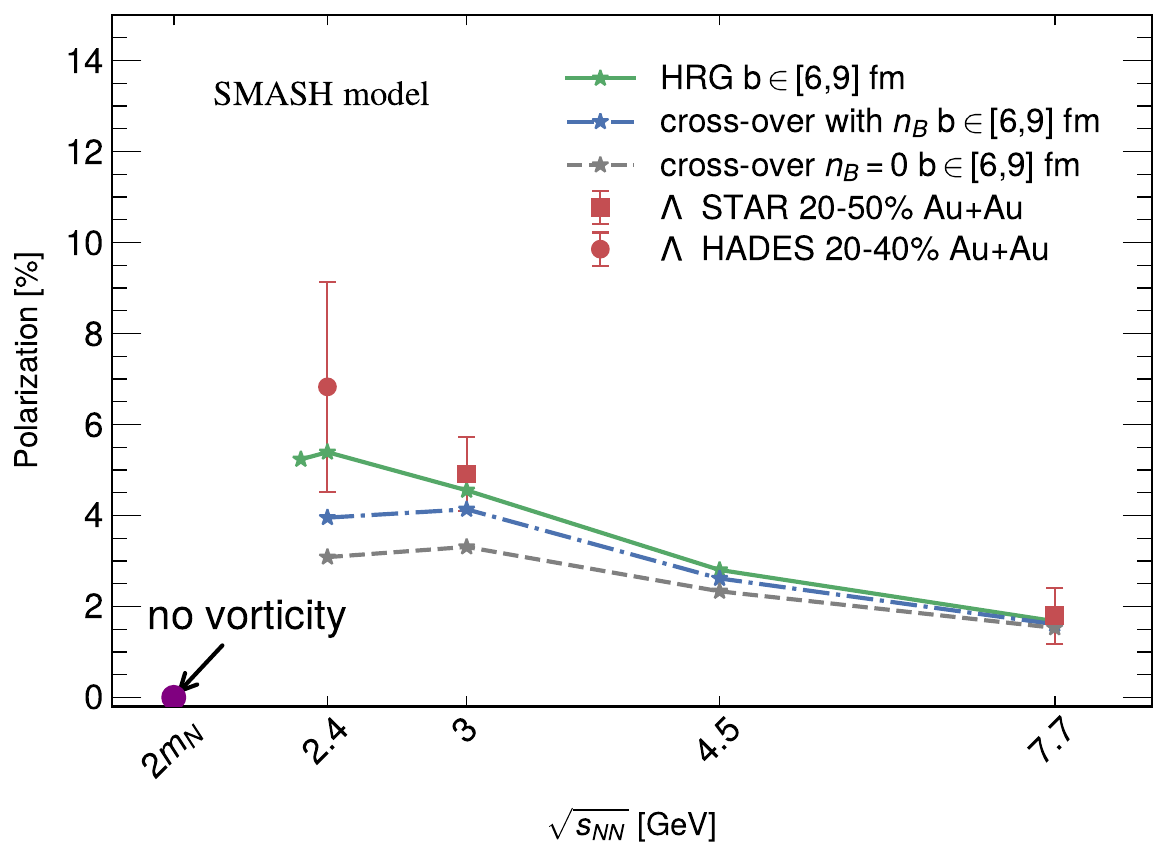}
\caption{
Global polarization of $\Lambda$ hyperons
as a function of collision energy in Au+Au collisions by
using the SMASH model. The gray dashed, blue dash-dotted, and green
solid lines represent the results with HotQCD, NEOS-BQS, and HRG EOS, respectively. The red
markers denote experimental data from the STAR~\cite{STAR:2023eck}
and HADES~\cite{HADES:2022enx} collaborations. The purple point corresponds to $\sqrt{s_{NN}} = 2m_{N}$, the threshold energy where no collisions occur, resulting in vanishing vorticity and polarization. The kinematic cuts
applied to $\Lambda$ hyperons are consistent with those used in the
experimental measurements:
$p_{T} \in [0.2, 1.5]$ GeV and $y_{cm} \in [-0.5, 0.3]$ at $\sqrt{s_{NN}} = 2.4$ GeV; 
$p_{T} \in [0.7, 2.0]$ GeV and $y_{cm} \in [-0.2, 1]$ for $2.4 < \sqrt{s_{NN}} < 7.7$ GeV; and 
$p_{T} \in [0.5, 6.0]$ GeV with $y_{cm} \in [-1, 1]$ at $\sqrt{s_{NN}} = 7.7$ GeV.
}
\label{fig:energy}
\end{figure*}


\subsection{EOS and temperature}
The local temperature can be extracted by using the EOS and  $\epsilon,n_{B}$.
We consider three kinds of EOS in this work.  We first validated our framework by reproducing existing results for the baryon-free crossover HotQCD EOS \cite{HotQCD:2014kol,HotQCD:2018pds}. 
Since the zero-baryon-density approximation is insufficient for describing low-energy collisions, we also employ two EOS that incorporate finite baryon density: the crossover NEOS-BQS EOS \cite{Monnai:2021kgu,McNelis:2021acu} and a purely hadronic EOS based on the Hadron Resonance Gas (HRG) model. Here, the NEOS-BQS EOS is derived by extending the HotQCD results to finite net baryon density using a Taylor expansion in the baryon chemical potential. 
 Since the proton is the lightest baryon in the HRG model, we emphasize that a finite temperature can only be extracted in the physical region where the energy density satisfies $\epsilon > m_{p}n_{B}$. 
Both of these EOS enforce strangeness neutrality and assume that the electric charge density $n_{Q}$ and net baryon density $n_{B}$ satisfy $n_{Q} = 0.4n_{B}$, where $0.4$ corresponds to the proton-to-nucleon ratio in gold (Au) nuclei.

To analyze the impact of different EOS, we compare the temperature as a function of energy density $\epsilon$ at fixed net baryon density $n_{B}$ for three EOS in Fig.~\ref{fig:EOS}. 
We observe that the NEOS-BQS EOS (blue dash-dotted lines) smoothly connects to the HotQCD EOS (gray dashed lines) at high temperature and overlaps with the HRG EOS (green solid lines) at low temperature. We note, however, that NEOS-BQS can still deviate from the HRG EOS at very low temperature due to the limitation of  Taylor expansion method, e.g. as shown in Fig.~\ref{fig:EOS}, differences become visible around $T\leq125$ MeV.

At high energy densities, we observe minimal differences in the extracted temperatures between the different EOS and various $n_{B}$ values. This indicates that temperature extraction is relatively insensitive to the EOS type and $n_{B}$ in this regime. This observation explains why similar global polarization are obtained using different methods at medium and high collision energies. 
While, as the collision energy decreases and the system enters the low-energy-density and high-$n_B$ regime, the temperature differences extracted from different EOS become increasingly significantly. By comparing HotQCD and NEOS-BQS EOS, we observe that including the effects of finite baryon density leads to a lower extracted temperature at the same energy density. 
Comparing the temperature from the NEOS-BQS and HRG EOS at the same energy density and baryon number density ($n_{B}$), we find that the temperature extracted from the HRG EOS is even lower. These differences highlight how finite baryon density and the state of matter (hadronic versus a mixed phase with QGP) significantly influence temperature extraction in low-energy collisions. Crucially, since the polarization pseudo vector $\mathcal{S}^\mu$ in Eq.~(\ref{eq:Smu}) scales inversely with temperature, this sensitivity of $T$ to the chosen EOS directly translates into an EOS dependence of the global polarization.

\subsection{Limitations of the current studies}
\label{subsec:limitations}
Before concluding this section, we would like to discuss the limitations of the current studies. In principle, $\overline{\Lambda}$ hyperons are also polarized in low-energy collisions. However, since the probability of generating $\overline{\Lambda}$ hyperons in low-energy collisions is very low, we cannot obtain a sufficient number of $\overline{\Lambda}$ hyperons even after running $10^5$ events. Therefore, in this study, we focus on the global polarization of $\Lambda$ hyperons. 

Additionally, we also study the global polarization of $\Lambda$ hyperons below the $\Lambda$ production threshold in nucleon-nucleon collisions. In such events, $\Lambda$ hyperons are generated through multiple scatterings. However, when $\sqrt{s_{NN}} < 2.3$ GeV, we cannot collect a sufficient number of $\Lambda$ hyperons to compute their global polarization due to the limitations of our computational resources. Therefore, we will only report our results for $\sqrt{s_{NN}} \geq 2.3$ GeV.

Finally, unlike in hydrodynamic models, the equation of state (EoS) is not a direct input in the transport approach employed for low beam energy collisions. Instead, we utilize only the temperature and fluid velocity that are extracted from the energy–momentum tensor under different equations of state. The EoS governing the evolution remains close to that of a hadron resonance gas. Currently, our method captures the most essential information (near freeze-out) through a phenomenological approach. In the future, we aim to develop a more self-consistent framework for incorporating the transport model under varying equations of state.

\section{Results and discussion}\label{sec:Results-and-discussion}
In this section, we present the results of global polarization for $\Lambda$ hyperons as functions of collision energy $\sqrt{s_{NN}}$, centrality, rapidity $y_{cm}$, and transverse momentum $p_T$ in Au+Au collisions at $\sqrt{s_{NN}} = 2.4$, $3$, $3.5$, $3.9$, $4.5$, and $7.7$ GeV using three different EOS.

\subsection{EOS and collision energy dependence}
\label{sec:EOS_dependence}

In Fig.~\ref{fig:energy}, we present EOS and collision energies dependence of  global polarization for $\Lambda$ hyperons  in 
$20 - 50\%$ Au+Au collisions.


The simulations with three different EOS, averaged
over impact parameters $b\in{[}6,9{]}$ fm, are shown
by three different lines. At $\sqrt{s_{NN}}=7.7$ GeV, all three theoretical
scenarios yield consistent global polarization that agree well
with the experimental data. However, as the collision energy decreases,
significant differences among the three EOS emerge. 
This result indicates that the global polarization in this regime is highly sensitive to the EOS.

The results using the HotQCD EOS, shown as the gray dashed line in Fig.~\ref{fig:energy}, are lower than the experimental data when $\sqrt{s_{NN}} < 7.7$ GeV and exhibit a non-monotonic trend—first increasing and then decreasing as the collision energy decreases—reaching a peak around $\sqrt{s_{NN}} = 3$ GeV. These results are similar to those obtained from the AMPT model in Ref.~\cite{Guo:2021udq}.


When the net baryon number effect is included, the global polarization of $\Lambda$ hyperons using the NEOS-BQS EOS, shown as the blue dash-dotted line in Fig.~\ref{fig:energy}, is found to be larger than that obtained with the HotQCD EOS. As shown in Fig.~\ref{fig:EOS}, for a fixed energy density, the temperature extracted from the NEOS-BQS EOS is lower than that from the HotQCD EOS.
According to the definition of thermal vorticity in Eq.~(\ref{eq:th vorticity}), a reduced temperature results in an enhanced thermal vorticity, which eventually amplifies the global polarization. 
Nevertheless, a maximum still appears around $\sqrt{s_{NN}} = 3$ GeV, and deviations from the experimental data persist at $\sqrt{s_{NN}} = 2.4$ GeV.


\begin{figure}[tbp]
\includegraphics[scale=0.4]{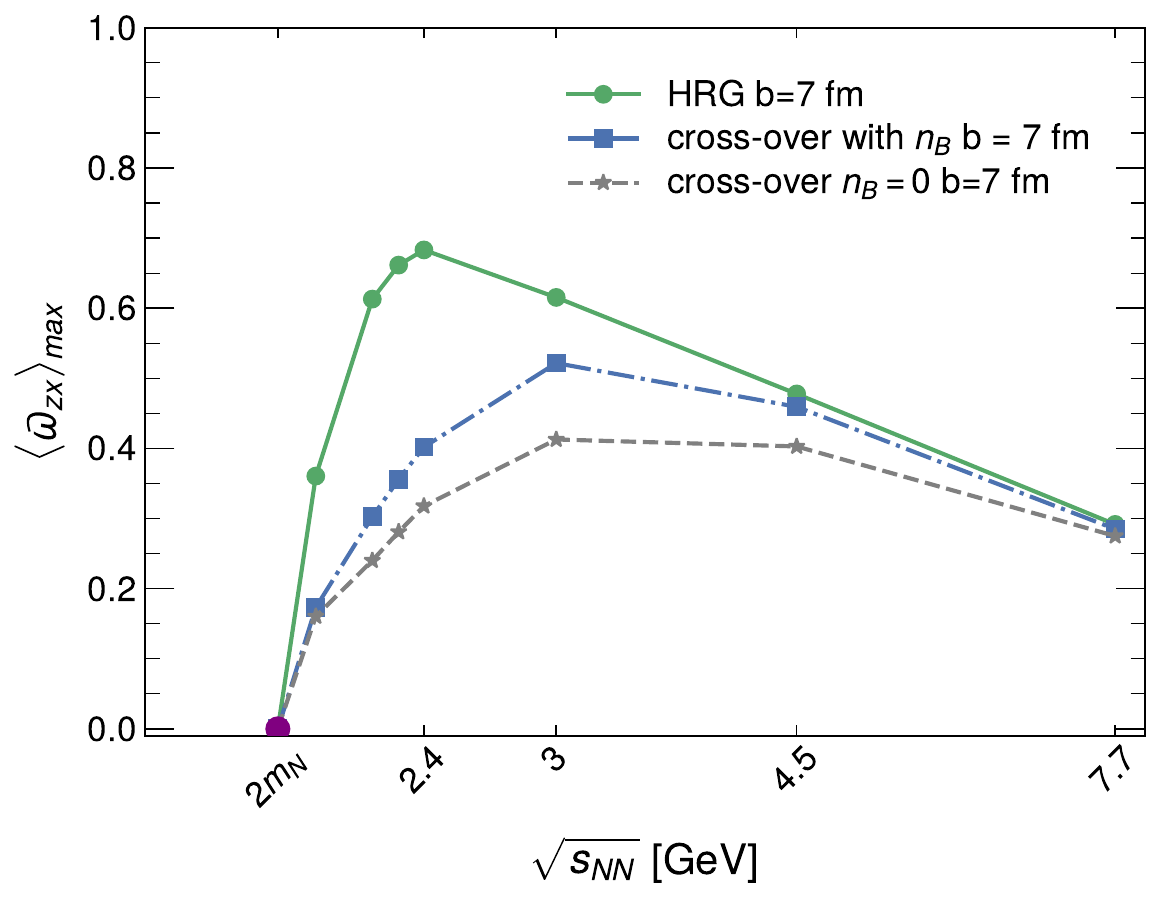}
\caption{The energy-weighted $\langle\varpi_{zx}\rangle_{max}$ component of thermal vorticity at space rapidity $\eta = 0$ as a function of collision energy $\sqrt{s_{NN}}$. The color assignment is the same as in Fig.~\ref{fig:energy}.}
\label{fig:vorticity}
\end{figure}

Remarkably, we observe that the results with the HRG EOS, shown as the green solid line in Fig.~\ref{fig:energy}, describe the experimental data well. The simulation with the HRG EOS provides the largest values compared to those obtained with the other two EOS. This is because the temperature extracted using the HRG EOS is smaller than that obtained with the HotQCD and NEOS-BQS EOS, as shown in Fig.~\ref{fig:EOS}. 
We also find that, with the HRG EOS, the global polarization of $\Lambda$ hyperons at $\sqrt{s_{NN}}=2.3$ GeV is slightly smaller than that at $\sqrt{s_{NN}}=2.4$ GeV.
Compared with the global polarization computed using the other two EOS, this indicates that the location of the global polarization peak is also sensitive to the EOS.

\subsection{Possible maximum global polarization}
As mentioned in the introduction, if the global polarization is induced by the thermal vorticity, it will vanish when $\sqrt{s_{NN}} \leq 2m_N$. 
On the other hand, we observe that the simulation with the HRG EOS does not exhibit a local maximum above $\sqrt{s_{NN}} = 2.4$ GeV, unlike the simulations using the other two EOS. This raises the question: where is the critical collision energy at which the maximum global polarization induced by thermal vorticity occurs?

\begin{figure}
\includegraphics[scale=0.43]{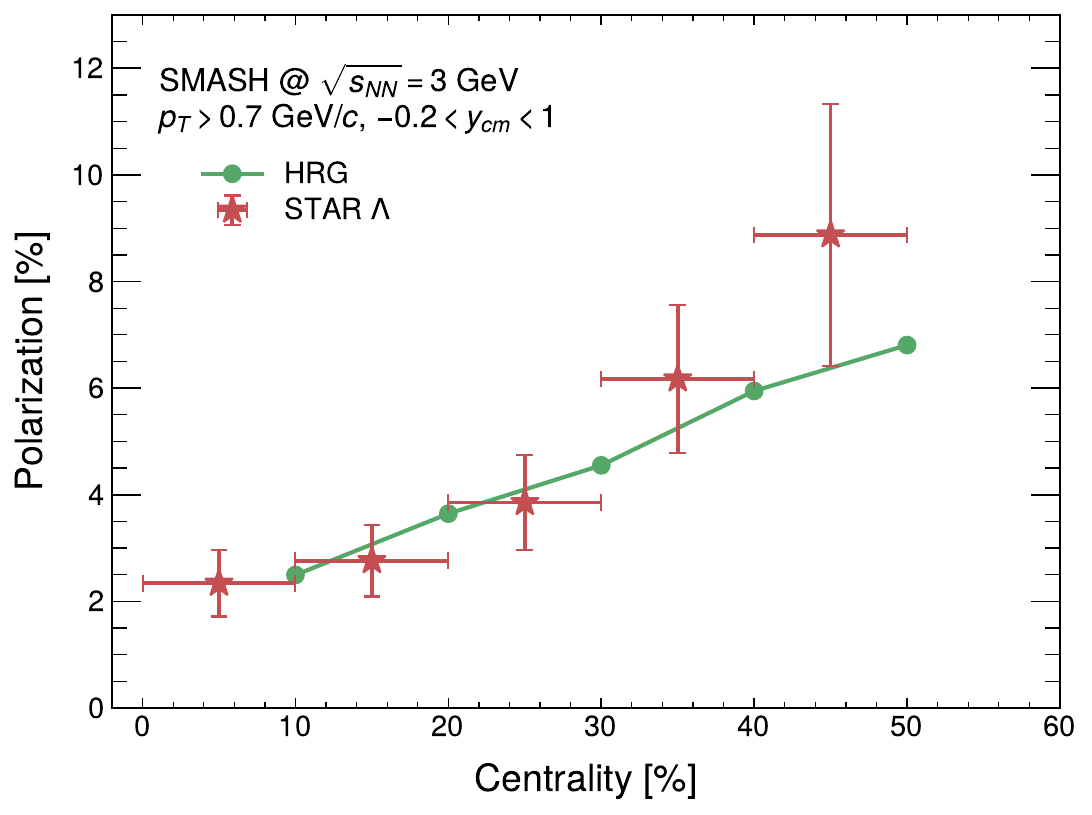}
\caption{Polarization of $\Lambda$ hyperons  along the $y$ direction as a function of centrality at $\sqrt{s_{NN}} = 3$ GeV in Au+Au collisions. The green solid line represents results computed using the HRG EOS. The red markers denote experimental data from the STAR~\cite{STAR:2023eck} collaboration. The kinematic region of $\Lambda$ hyperons is the same as that in Fig.~\ref{fig:energy}.}
\label{fig:centrality}
\end{figure}

Unfortunately, it is challenging to simulate global polarization when $\sqrt{s_{NN}} < 2.3$ GeV. As mentioned in Sec.~\ref{subsec:limitations}, $\sqrt{s_{NN}} = 2.3$ GeV is already below the $\Lambda$ production threshold in nucleon-nucleon collisions, and the particle yield is significantly suppressed. Consequently, we cannot collect a sufficient number of $\Lambda$ hyperons generated from multiple scatterings when $\sqrt{s_{NN}} < 2.3$ GeV.

However, given that the global polarization along the $y$-direction is strongly positively correlated with the thermal vorticity component $\varpi_{zx}$, we estimate the possible maximum value of global polarization by studying the system's thermal vorticity. 
In Fig.~\ref{fig:vorticity}, we show the collision energy dependence of the energy-weighted maximum thermal-vorticity component, $\langle\varpi_{zx}\rangle_{max}$. Following Ref.~\cite{Deng:2020ygd}, we average $\langle\varpi_{zx}\rangle$ over the transverse plane in the midrapidity region ($\eta = 0$) at $b = 7$ fm. The $\langle\varpi_{zx}\rangle$ first increases and then decreases with time; for each event, we record only its maximum value. The $\langle\varpi_{zx}\rangle_{max}$ serves as a proxy for the magnitude of the deposited thermal vorticity.
It is instructive to note that in the low-energy regime, the well-known squeeze-out effect preferentially drives particle emission out of the reaction plane \citep{Gutbrod:1989gh,Bass:1993ce,Hartnack:1994ce}. This mechanism not only gives rise to a negative elliptic flow but also distorts the local vorticity field structure. From a physical perspective, the squeeze-out dynamics are expected to exert a profound influence on the local longitudinal  vorticity along the beam direction, but may have a limited impact on the arveraged global vorticty  $\langle\varpi_{zx}\rangle$ .  Therefore, as the collision energy decreases, this  $\langle\varpi_{zx}\rangle_{max}$  still exhibits a smooth trend of rising and then falling. As shown in Fig. \ref{fig:vorticity}. The  $\langle\varpi_{zx}\rangle_{max}$ calculated using the HRG EOS peaks around $\sqrt{s_{NN}} = 2.4$ GeV, whereas the results from other two EOS reach their maximum around $\sqrt{s_{NN}} = 3$ GeV. 
This finding implies that the peak of the global polarization for $\Lambda$ hyperons may also located at $\sqrt{s_{NN}} = 2.4$ GeV for HRG EOS according to Eq.~(\ref{eq:Smu}).



\subsection{Centrality, rapidity and $p_T$ dependence at $\sqrt{s_{NN}}=3$ GeV}

The STAR collaboration \cite{STAR:2023eck} also measured the centrality, rapidity, and $p_T$ dependence of polarization along $-y$ direction, $-P_y$, for $\Lambda$ hyperons at $\sqrt{s_{NN}} = 3$ GeV in Au+Au collisions. 
Given that our investigation into EOS sensitivity in Sec.~\ref{sec:EOS_dependence} identified the HRG EOS as the only scenario capable of reproducing the magnitude of the experimental data, we will use the HRG EOS exclusively for the following analysis.

We present the centrality dependence of $-P_y$ in $20$–$50\%$ Au+Au collisions at $\sqrt{s_{NN}} = 3$ GeV in Fig.~\ref{fig:centrality}. Our results describe the experimental data well.
Additionally, we observe that $-P_y$ increases as the collision becomes more peripheral (i.e., as centrality increases). This centrality dependence is attributed to the combined effects of a lower freeze-out temperature and increased kinematic vorticity, due to the stronger geometric anisotropy in peripheral collisions.

\begin{figure}
\includegraphics[scale=0.43]{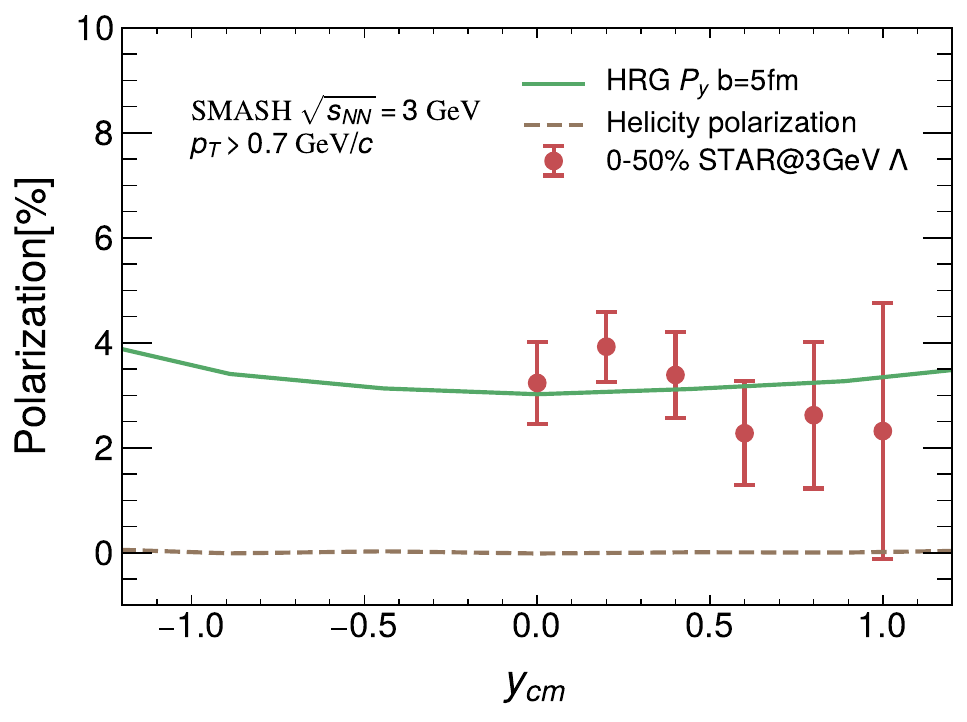}\caption{Polarization of $\Lambda$ hyperons along the $-y$ direction $-P_y$ and helicity polarization of $\Lambda$ hyperons as a function of rapidity $y_{cm}$ in $0$–$50\%$ centrality Au+Au collisions at $\sqrt{s_{NN}} = 3$ GeV. The green solid line and brown dashed line represent results for $-P_y$ and helicity polarization, respectively. The red markers denote experimental data from the STAR collaboration~\cite{STAR:2023eck}. We have chosen the HRG EOS at $b = 5$ fm and take $p_T > 0.7$ GeV for $\Lambda$ hyperons.
}\label{fig:rapidity}
\end{figure}

\begin{figure}
\includegraphics[scale=0.43]{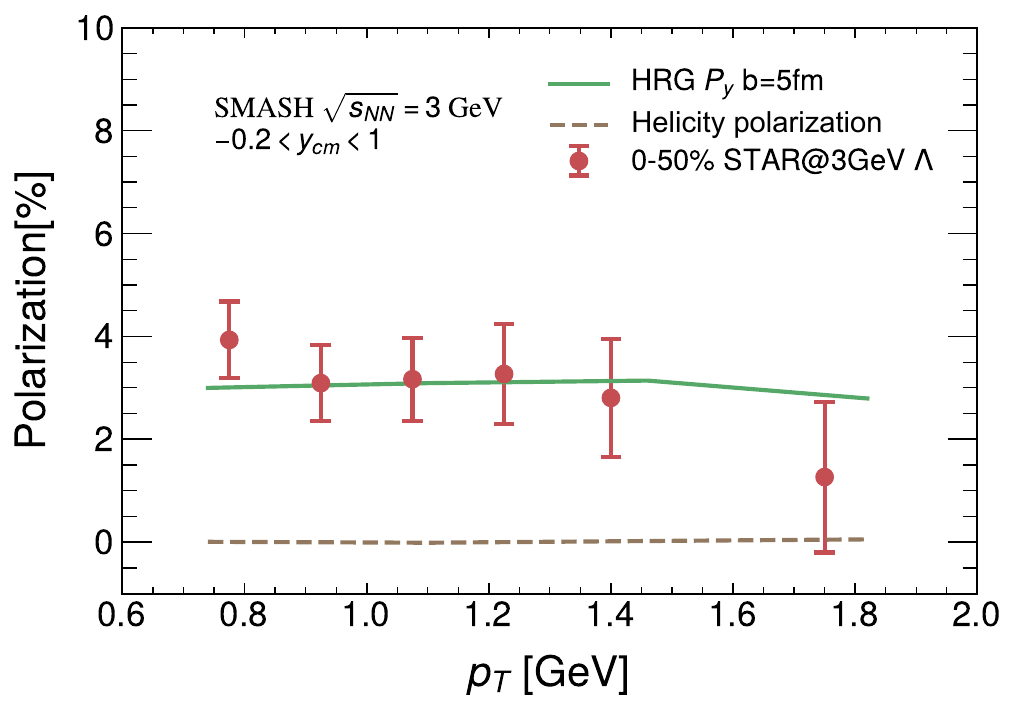}\caption{
Polarization of $\Lambda$ hyperons along the $-y$ direction $-P_y$ and helicity polarization of $\Lambda$ hyperons as a function of transverse momentum $p_T$ in $0$–$50\%$ centrality Au+Au collisions at $\sqrt{s_{NN}} = 3$ GeV. The green solid line and brown dashed line represent results for $-P_y$ and helicity polarization, respectively. The red markers denote experimental data from the STAR collaboration~\cite{STAR:2023eck}. We have chosen the HRG EOS at $b = 5$ fm and take $-0.2 < y_{cm} < 1$ for $\Lambda$ hyperons.
}\label{fig:pT}
\end{figure}

The polarization along the $y$-direction of $\Lambda$ hyperons as functions of rapidity $y_{cm}$ and transverse momentum $p_T$ in $0$–$50\%$ centrality Au+Au collisions at $\sqrt{s_{NN}} = 3$ GeV are shown in Fig.~\ref{fig:rapidity} and Fig.~\ref{fig:pT}, respectively. We find that our simulations agree with the experimental data. For the rapidity dependence, we observe a slight increase in $-P_y$ with increasing $|y_{cm}|$, whereas no significant dependence on $p_T$ is observed within the examined range.


\section{Brief discussion on helicity polarization}
\label{sec:helicity}
Helicity polarization ~\citep{Becattini:2020xbh,Gao:2021rom,Yi:2021unq,Yi:2023tgg} is another interesting quantity that reflects information for polarization. For example, the helicity polarization as a function of the azimuthal angle can provide additional insights into the fine structure of kinetic vorticity ~\citep{Yi:2023tgg}.
The helicity polarization for a relativistic particle is defined as:
\begin{eqnarray}
S^{h} & = & \frac{\mathbf{p}\cdot \mathcal{\boldsymbol{S}}(\mathbf{p})}{|\mathbf{p}|},\label{eq:def_Sh}
\end{eqnarray}
where $\mathbf{p}$ is the particle momentum, $\boldsymbol{\mathcal{S}}(\mathbf{p})$ is the polarization pesudo vector.

Here, we focus on the helicity polarization of $\Lambda$ hyperons induced by the thermal vorticity or other collective motions of the system, namely $S_{\textrm{hydro}}^{h}$. 
It has been found that the azimuthal-angle dependence of the helicity polarization is mainly driven by thermal vorticity in low energies collisions \cite{Yi:2021unq,Yi:2023tgg}. 
In this case, it has been pointed out in Refs.~\citep{Becattini:2020xbh,Gao:2021rom,Yi:2021unq,Yi:2023tgg} that, in the absence of axial charge, the system exhibits space-reversal symmetry, which leads to
\begin{equation}
    S_{\textrm{hydro}}^{h}(\mathbf{p})=-S_{\textrm{hydro}}^{h}(-\mathbf{p}).
\end{equation}
As a result, after integrating over azimuthal angle, $S_{\textrm{hydro}}^{h}$ exhibits rapidity and $p_T$ dependence, as shown by the brown dashed lines in Figs.~\ref{fig:rapidity} and~\ref{fig:pT}.

Therefore, if a finite rapidity- or $p_T$-dependent helicity polarization is measured in experiments, such a non-vanishing helicity polarization cannot be attributed to thermal vorticity or other collective motions of the system. On the other hand, we emphasize that helicity polarization can also be induced by other QCD effects, such as polarized fragmentation functions see e.g. Refs.~\citep{deFlorian:1997zj,Callos:2020qtu,DAlesio:2020wjq,Chen:2021zrr,Chen:2021hdn} and references therein.

\section{Summary and discussion}
In this work, we have studied the global polarization of $\Lambda$ hyperons in low-energy collisions using the transport model SMASH with three different EOS: HotQCD, NEOS-BQS, and HRG EOS.

Remarkably, we have observed that the global polarization at low collision energies is sensitive to EOS and only the simulations with the HRG EOS describe the experimental data well, as shown in Fig.~\ref{fig:energy}. These results indicate that the global polarization of $\Lambda$ hyperons can still be explained by thermal vorticity 
of hadronic matters
in low-energy collisions, even when it is below 
$\Lambda$ hyperon production threshold in nucleon-nucleon collisions. The sensitivity to the choice of EOS highlights the importance of properly accounting for finite baryon density and the hadronic nature of matter at low collision energies.

Furthermore, the estimation of the energy-weighted maximum thermal vorticity component, $\langle\varpi_{zx}\rangle_{max}$, computed with HRG EOS,
suggests that a possible polarization maximum occurs at $\sqrt{s_{NN}} = 2.4$ GeV, which is lower than 
the results by using other two EOS and
previously predictions from the models that neglect baryon density effects. This also indicates that the location of the global polarization peak is sensitive to the EOS.

We have also studied the centrality, rapidity, and $p_T$ dependence of $-P_y$ at $\sqrt{s_{NN}} = 3$ GeV in Au+Au collisions. Our simulations with the HRG EOS also show good agreement with the experimental data.



Finally, we briefly discuss helicity polarization. Due to the system's space-reversal symmetry, after azimuthal angle integration, the rapidity- and $p_T$-dependent helicity polarization induced by thermal vorticity or other collective motions of the system vanishes.

Before ending this work, we would like to comment on our assumption of equilibrium. In principle, the system created in low-energy collisions may not fully reach thermal global or local equilibrium. There could be additional out-of-equilibrium corrections to the modified Cooper-Frye formula. Furthermore, the $\Lambda$ production mechanism is expected to be different at low collision energies. Particularly in the subthreshold regime, $\Lambda$ hyperons are mainly generated through multiple hadronic scatterings, and a fraction of them may originate from regions outside the nucleus overlap zone, i.e., the spectator region. In this situation, the fluid vorticity picture will break down. These effects are left for future investigations.

\section*{Acknowledgment}
This work is supported in part by National Key Research and Development Program of China under Contract No. 2022YFA1605500, by National Natural Science Foundation of China (NSFC) under Grants No.~12225503, 12135011, 12547164, 12535010, 12435009
by the Chinese Academy of Sciences (CAS) under Grant No. YSBR-088, supported
by the China Postdoctoral Science Foundation under Grant Number 2025M783388.

\bibliographystyle{h-physrev}
\bibliography{qkt-ref-2}

@article{Chen:2021zrr,
    author = "Chen, Kai-bao and Liang, Zuo-tang and Song, Yu-kun and Wei, Shu-yi",
    title = "{Longitudinal and transverse polarizations of {\ensuremath{\Lambda}} hyperon in unpolarized SIDIS and e+e- annihilation}",
    eprint = "2108.07740",
    archivePrefix = "arXiv",
    primaryClass = "hep-ph",
    doi = "10.1103/PhysRevD.105.034027",
    journal = "Phys. Rev. D",
    volume = "105",
    number = "3",
    pages = "034027",
    year = "2022"
}

@article{deFlorian:1997zj,
    author = "de Florian, D. and Stratmann, M. and Vogelsang, W.",
    title = "{QCD analysis of unpolarized and polarized Lambda baryon production in leading and next-to-leading order}",
    eprint = "hep-ph/9711387",
    archivePrefix = "arXiv",
    reportNumber = "CERN-TH-97-286, DTP-97-106",
    doi = "10.1103/PhysRevD.57.5811",
    journal = "Phys. Rev. D",
    volume = "57",
    pages = "5811--5824",
    year = "1998"
}

@article{Callos:2020qtu,
    author = "Callos, Daniel and Kang, Zhong-Bo and Terry, John",
    title = "{Extracting the transverse momentum dependent polarizing fragmentation functions}",
    eprint = "2003.04828",
    archivePrefix = "arXiv",
    primaryClass = "hep-ph",
    doi = "10.1103/PhysRevD.102.096007",
    journal = "Phys. Rev. D",
    volume = "102",
    number = "9",
    pages = "096007",
    year = "2020"
}

@article{DAlesio:2020wjq,
    author = "D'Alesio, Umberto and Murgia, Francesco and Zaccheddu, Marco",
    title = "{First extraction of the $\Lambda$ polarizing fragmentation function from Belle $e^+e^-$ data}",
    eprint = "2003.01128",
    archivePrefix = "arXiv",
    primaryClass = "hep-ph",
    doi = "10.1103/PhysRevD.102.054001",
    journal = "Phys. Rev. D",
    volume = "102",
    number = "5",
    pages = "054001",
    year = "2020"
}

@article{Chen:2021hdn,
    author = "Chen, Kai-Bao and Liang, Zuo-Tang and Pan, Yan-Lei and Song, Yu-Kun and Wei, Shu-Yi",
    title = "{Isospin Symmetry of Fragmentation Functions}",
    eprint = "2102.00658",
    archivePrefix = "arXiv",
    primaryClass = "hep-ph",
    doi = "10.1016/j.physletb.2021.136217",
    journal = "Phys. Lett. B",
    volume = "816",
    pages = "136217",
    year = "2021"
}

@article{Yamamoto:2023okm,
    author = "Yamamoto, Naoki and Yang, Di-Lun",
    title = "{Chiral kinetic theory with self-energy corrections and neutrino spin Hall effect}",
    eprint = "2308.08257",
    archivePrefix = "arXiv",
    primaryClass = "hep-ph",
    doi = "10.1103/PhysRevD.109.056010",
    journal = "Phys. Rev. D",
    volume = "109",
    number = "5",
    pages = "056010",
    year = "2024"
}

@article{Lin:2022tma,
    author = "Lin, Shu and Wang, Ziyue",
    title = "{Shear induced polarization: collisional contributions}",
    eprint = "2206.12573",
    archivePrefix = "arXiv",
    primaryClass = "hep-ph",
    doi = "10.1007/JHEP12(2022)030",
    journal = "JHEP",
    volume = "12",
    pages = "030",
    year = "2022"
}

@article{Wang:2022yli,
    author = "Wang, Ziyue",
    title = "{Spin evolution of massive fermion in QED plasma}",
    eprint = "2205.09334",
    archivePrefix = "arXiv",
    primaryClass = "hep-ph",
    doi = "10.1103/PhysRevD.106.076011",
    journal = "Phys. Rev. D",
    volume = "106",
    number = "7",
    pages = "076011",
    year = "2022"
}

@article{Fang:2025pzy,
    author = "Fang, Shuo and Pu, Shi and Yang, Di-Lun",
    title = "{Radiative corrections on vortical spin polarization in hot QCD matter}",
    eprint = "2503.13320",
    archivePrefix = "arXiv",
    primaryClass = "hep-ph",
    doi = "10.1103/tds1-793m",
    journal = "Phys. Rev. D",
    volume = "112",
    number = "1",
    pages = "014038",
    year = "2025"
}

@article{Wang:2025mfz,
    author = "Wang, Jia-Rong and Fang, Shuo and Yang, Di-Lun and Pu, Shi",
    title = "{Is the shear induced spin polarization non-dissipative?}",
    eprint = "2507.15238",
    archivePrefix = "arXiv",
    primaryClass = "hep-ph",
    month = "7",
    year = "2025"
}

@article{ALICE:2019onw,
    author = "Acharya, Shreyasi and others",
    collaboration = "ALICE",
    title = "{Global polarization of $\Lambda \bar \Lambda$ hyperons in Pb-Pb collisions at $\sqrt {s_{NN}}$ = 2.76 and 5.02 TeV}",
    eprint = "1909.01281",
    archivePrefix = "arXiv",
    primaryClass = "nucl-ex",
    reportNumber = "CERN-EP-2019-173",
    doi = "10.1103/PhysRevC.101.044611",
    journal = "Phys. Rev. C",
    volume = "101",
    number = "4",
    pages = "044611",
    year = "2020",
    note = "[Erratum: Phys.Rev.C 105, 029902 (2022)]"
}

@article{Palermo:2022lvh,
    author = "Palermo, Andrea and Becattini, Francesco and Buzzegoli, Matteo and Inghirami, Gabriele and Karpenko, Iurii",
    title = "{Local equilibrium and Lambda polarization in high energy heavy ion collisions}",
    eprint = "2208.09874",
    archivePrefix = "arXiv",
    primaryClass = "nucl-th",
    doi = "10.1051/epjconf/202327601026",
    journal = "EPJ Web Conf.",
    volume = "276",
    pages = "01026",
    year = "2023"
}

@article{Buzzegoli:2022fxu,
    author = "Buzzegoli, M. and Becattini, F. and Inghirami, G. and Karpenko, I. and Palermo, A.",
    title = "{Spin-thermal Shear Coupling in Relativistic Nuclear Collisions}",
    eprint = "2208.04449",
    archivePrefix = "arXiv",
    primaryClass = "nucl-th",
    doi = "10.5506/APhysPolBSupp.16.1-A39",
    journal = "Acta Phys. Polon. Supp.",
    volume = "16",
    number = "1",
    pages = "39",
    year = "2023"
}

@article{Becattini:2022zvf,
    author = "Becattini, Francesco",
    title = "{Spin and polarization: a new direction in relativistic heavy ion physics}",
    eprint = "2204.01144",
    archivePrefix = "arXiv",
    primaryClass = "nucl-th",
    doi = "10.1088/1361-6633/ac97a9",
    journal = "Rept. Prog. Phys.",
    volume = "85",
    number = "12",
    pages = "122301",
    year = "2022"
}

@article{Wu:2022mkr,
    author = "Wu, Xiang-Yu and Yi, Cong and Qin, Guang-You and Pu, Shi",
    title = "{Local and global polarization of \ensuremath{\Lambda} hyperons across RHIC-BES energies: The roles of spin hall effect, initial condition, and baryon diffusion}",
    eprint = "2204.02218",
    archivePrefix = "arXiv",
    primaryClass = "hep-ph",
    doi = "10.1103/PhysRevC.105.064909",
    journal = "Phys. Rev. C",
    volume = "105",
    number = "6",
    pages = "064909",
    year = "2022"
}

@article{Florkowski:2021xvy,
    author = "Florkowski, Wojciech and Kumar, Avdhesh and Mazeliauskas, Aleksas and Ryblewski, Radoslaw",
    title = "{Effect of thermal shear on longitudinal spin polarization in a thermal model}",
    eprint = "2112.02799",
    archivePrefix = "arXiv",
    primaryClass = "hep-ph",
    doi = "10.1103/PhysRevC.105.064901",
    journal = "Phys. Rev. C",
    volume = "105",
    number = "6",
    pages = "064901",
    year = "2022"
}

@article{Becattini:2021suc,
    author = "Becattini, F. and Buzzegoli, M. and Palermo, A.",
    title = "{Spin-thermal shear coupling in a relativistic fluid}",
    eprint = "2103.10917",
    archivePrefix = "arXiv",
    primaryClass = "nucl-th",
    doi = "10.1016/j.physletb.2021.136519",
    journal = "Phys. Lett. B",
    volume = "820",
    pages = "136519",
    year = "2021"
}

@article{STAR:2018gyt,
    author = "Adam, Jaroslav and others",
    collaboration = "STAR",
    title = "{Global polarization of $\Lambda$ hyperons in Au+Au collisions at $\sqrt{s_{_{NN}}}$ = 200 GeV}",
    eprint = "1805.04400",
    archivePrefix = "arXiv",
    primaryClass = "nucl-ex",
    doi = "10.1103/PhysRevC.98.014910",
    journal = "Phys. Rev. C",
    volume = "98",
    pages = "014910",
    year = "2018"
}

@Article{STAR:2017ckg,
  Title                    = {{Global $\Lambda$ hyperon polarization in nuclear collisions: evidence for the most vortical fluid}},
  Author                   = {Adamczyk, L. and others},
  DOI                      = {10.1038/nature23004},
  Eprint                   = {1701.06657},
  Pages                    = {62-65},
  Volume                   = {548},
  Year                     = {2017},
  Archiveprefix            = {arXiv},
  Collaboration            = {STAR},
  Journal                  = {Nature},
  Primaryclass             = {nucl-ex},
  Slaccitation             = {%%CITATION = ARXIV:1701.06657;%%}
}

@Article{Becattini:2013fla,
  Title                    = {{Relativistic distribution function for particles with spin at local thermodynamical equilibrium}},
  Author                   = {Becattini, F. and Chandra, V. and Del Zanna, L. and Grossi, E.},
  DOI                      = {10.1016/j.aop.2013.07.004},
  Eprint                   = {1303.3431},
  Pages                    = {32--49},
  Volume                   = {338},
  Year                     = {2013},
  Archiveprefix            = {arXiv},
  Journal                  = {Annals Phys.},
  Primaryclass             = {nucl-th}
}

@Article{Becattini:2013vja,
  Title                    = {{$\Lambda$ polarization in peripheral heavy ion collisions}},
  Author                   = {Becattini, F. and Csernai, L. and Wang, D.J.},
  DOI                      = {10.1103/PhysRevC.88.034905},
  Eprint                   = {1304.4427},
  Note                     = {[Erratum: Phys.Rev.C 93, 069901 (2016)]},
  Number                   = {3},
  Pages                    = {034905},
  Volume                   = {88},
  Year                     = {2013},
  Archiveprefix            = {arXiv},
  Journal                  = {Phys. Rev. C},
  Primaryclass             = {nucl-th}
}

@Article{Becattini:2007nd,
  Title                    = {{The Ideal relativistic spinning gas: Polarization and spectra}},
  Author                   = {Becattini, F. and Piccinini, F.},
  DOI                      = {10.1016/j.aop.2008.01.001},
  Eprint                   = {0710.5694},
  Pages                    = {2452--2473},
  Volume                   = {323},
  Year                     = {2008},
  Archiveprefix            = {arXiv},
  Journal                  = {Annals Phys.},
  Primaryclass             = {nucl-th}
}

@Article{Betz:2007kg,
  Title                    = {{Polarization probes of vorticity in heavy ion collisions}},
  Author                   = {Betz, Barbara and Gyulassy, Miklos and Torrieri, Giorgio},
  DOI                      = {10.1103/PhysRevC.76.044901},
  Eprint                   = {0708.0035},
  Pages                    = {044901},
  Volume                   = {76},
  Year                     = {2007},
  Archiveprefix            = {arXiv},
  Journal                  = {Phys. Rev. C},
  Primaryclass             = {nucl-th}
}

@Article{Csernai:2013bqa,
  Title                    = {{Flow Vorticity in Peripheral High Energy Heavy Ion Collisions}},
  Author                   = {Csernai, L.P. and Magas, V.K. and Wang, D.J.},
  DOI                      = {10.1103/PhysRevC.87.034906},
  Eprint                   = {1302.5310},
  Number                   = {3},
  Pages                    = {034906},
  Volume                   = {87},
  Year                     = {2013},
  Archiveprefix            = {arXiv},
  Journal                  = {Phys. Rev. C},
  Primaryclass             = {nucl-th}
}

@Article{Gao:2020vbh,
  Title                    = {{Recent developments in chiral and spin polarization effects in heavy-ion collisions}},
  Author                   = {Gao, Jian-Hua and Ma, Guo-Liang and Pu, Shi and Wang, Qun},
  DOI                      = {10.1007/s41365-020-00801-x},
  Eprint                   = {2005.10432},
  Number                   = {9},
  Pages                    = {90},
  Volume                   = {31},
  Year                     = {2020},
  Archiveprefix            = {arXiv},
  Journal                  = {Nucl. Sci. Tech.},
  Primaryclass             = {hep-ph}
}

@Article{Hidaka:2017auj,
  Title                    = {{Nonlinear Responses of Chiral Fluids from Kinetic Theory}},
  Author                   = {Hidaka, Yoshimasa and Pu, Shi and Yang, Di-Lun},
  DOI                      = {10.1103/PhysRevD.97.016004},
  Eprint                   = {1710.00278},
  Number                   = {1},
  Pages                    = {016004},
  Volume                   = {D97},
  Year                     = {2018},
  Archiveprefix            = {arXiv},
  Journal                  = {Phys. Rev.},
  Primaryclass             = {hep-th},
  Reportnumber             = {RIKEN-QHP-260, RIKEN-iTHEMS-Report-17},
  Slaccitation             = {%%CITATION = ARXIV:1710.00278;%%}
}

@article{Karpenko:2016jyx,
    author = "Karpenko, I. and Becattini, F.",
    title = "{Study of $\Lambda $ polarization in relativistic nuclear collisions at $\sqrt{s_\mathrm {NN}}=7.7$ \textendash{}200 GeV}",
    eprint = "1610.04717",
    archivePrefix = "arXiv",
    primaryClass = "nucl-th",
    doi = "10.1140/epjc/s10052-017-4765-1",
    journal = "Eur. Phys. J. C",
    volume = "77",
    number = "4",
    pages = "213",
    year = "2017"
}

@article{Shi:2017wpk,
    author = "Shi, Shuzhe and Li, Kangle and Liao, Jinfeng",
    title = "{Searching for the Subatomic Swirls in the CuCu and CuAu Collisions}",
    eprint = "1712.00878",
    archivePrefix = "arXiv",
    primaryClass = "nucl-th",
    doi = "10.1016/j.physletb.2018.09.066",
    journal = "Phys. Lett. B",
    volume = "788",
    pages = "409--413",
    year = "2019"
}

@Article{Shi:2019wzi,
  Title                    = {{Signatures of Chiral Magnetic Effect in the Collisions of Isobars}},
  Author                   = {Shi, Shuzhe and Zhang, Hui and Hou, Defu and Liao, Jinfeng},
  Eprint                   = {1910.14010},
  Year                     = {2019},
  Archiveprefix            = {arXiv},
  Primaryclass             = {nucl-th},
  Slaccitation             = {%%CITATION = ARXIV:1910.14010;%%}
}

@Article{Sun:2017xhx,
  Title                    = {{$\Lambda$ hyperon polarization in relativistic heavy ion collisions from a chiral kinetic approach}},
  Author                   = {Sun, Yifeng and Ko, Che Ming},
  DOI                      = {10.1103/PhysRevC.96.024906},
  Eprint                   = {1706.09467},
  Issue                    = {2},
  Month                    = {Aug},
  Number                   = {2},
  Pages                    = {024906},
  URL                      = {https://link.aps.org/doi/10.1103/PhysRevC.96.024906},
  Volume                   = {C96},
  Year                     = {2017},
  Archiveprefix            = {arXiv},
  Journal                  = {Phys. Rev.},
  Numpages                 = {6},
  Primaryclass             = {nucl-th},
  Publisher                = {American Physical Society},
  Slaccitation             = {%%CITATION = ARXIV:1706.09467;%%}
}

@Article{Xia:2018tes,
  Title                    = {{Probing vorticity structure in heavy-ion collisions by local $\Lambda$ polarization}},
  Author                   = {Xia, Xiao-Liang and Li, Hui and Tang, Ze-Bo and Wang, Qun},
  DOI                      = {10.1103/PhysRevC.98.024905},
  Eprint                   = {1803.00867},
  Pages                    = {024905},
  Volume                   = {98},
  Year                     = {2018},
  Archiveprefix            = {arXiv},
  Journal                  = {Phys. Rev. C},
  Primaryclass             = {nucl-th}
}

@article{Xie:2017upb,
    author = "Xie, Yilong and Wang, Dujuan and Csernai, L\'aszl\'o P.",
    title = "{Global \ensuremath{\Lambda} polarization in high energy collisions}",
    eprint = "1703.03770",
    archivePrefix = "arXiv",
    primaryClass = "nucl-th",
    doi = "10.1103/PhysRevC.95.031901",
    journal = "Phys. Rev. C",
    volume = "95",
    number = "3",
    pages = "031901",
    year = "2017"
}

@article{Fu:2021pok,
    author = "Fu, Baochi and Liu, Shuai Y. F. and Pang, Longgang and Song, Huichao and Yin, Yi",
    title = "{Shear-Induced Spin Polarization in Heavy-Ion Collisions}",
    eprint = "2103.10403",
    archivePrefix = "arXiv",
    primaryClass = "hep-ph",
    doi = "10.1103/PhysRevLett.127.142301",
    journal = "Phys. Rev. Lett.",
    volume = "127",
    number = "14",
    pages = "142301",
    year = "2021"
}

@Article{Fu:2020oxj,
  author        = {Fu, Baochi and Xu, Kai and Huang, Xu-Guang and Song, Huichao},
  journal       = {Phys. Rev. C},
  title         = {{Hydrodynamic study of hyperon spin polarization in relativistic heavy ion collisions}},
  year          = {2021},
  number        = {2},
  pages         = {024903},
  volume        = {103},
  archiveprefix = {arXiv},
  doi           = {10.1103/PhysRevC.103.024903},
  eprint        = {2011.03740},
  primaryclass  = {nucl-th},
}

@article{Becattini:2021iol,
    author = "Becattini, F. and Buzzegoli, M. and Inghirami, G. and Karpenko, I. and Palermo, A.",
    title = "{Local Polarization and Isothermal Local Equilibrium in Relativistic Heavy Ion Collisions}",
    eprint = "2103.14621",
    archivePrefix = "arXiv",
    primaryClass = "nucl-th",
    doi = "10.1103/PhysRevLett.127.272302",
    journal = "Phys. Rev. Lett.",
    volume = "127",
    number = "27",
    pages = "272302",
    year = "2021"
}

@article{Liu:2021uhn,
	author = "Liu, Shuai Y. F. and Yin, Yi",
	title = "{Spin polarization induced by the hydrodynamic gradients}",
	eprint = "2103.09200",
	archivePrefix = "arXiv",
	primaryClass = "hep-ph",
	doi = "10.1007/JHEP07(2021)188",
	journal = "JHEP",
	volume = "07",
	pages = "188",
	year = "2021"
}

@article{Liu:2020dxg,
	author = "Liu, Shuai Y. F. and Yin, Yi",
	title = "{Spin Hall effect in heavy-ion collisions}",
	eprint = "2006.12421",
	archivePrefix = "arXiv",
	primaryClass = "nucl-th",
	doi = "10.1103/PhysRevD.104.054043",
	journal = "Phys. Rev. D",
	volume = "104",
	number = "5",
	pages = "054043",
	year = "2021"
}

@article{Gao:2021rom,
    author = "Gao, Jian-Hua",
    title = "{Helicity polarization in relativistic heavy ion collisions}",
    eprint = "2105.08293",
    archivePrefix = "arXiv",
    primaryClass = "hep-ph",
    doi = "10.1103/PhysRevD.104.076016",
    journal = "Phys. Rev. D",
    volume = "104",
    number = "7",
    pages = "076016",
    year = "2021"
}

@article{Yi:2021ryh,
	author = "Yi, Cong and Pu, Shi and Yang, Di-Lun",
	title = "{Reexamination of local spin polarization beyond global equilibrium in relativistic heavy ion collisions}",
	eprint = "2106.00238",
	archivePrefix = "arXiv",
	primaryClass = "hep-ph",
	doi = "10.1103/PhysRevC.104.064901",
	journal = "Phys. Rev. C",
	volume = "104",
	number = "6",
	pages = "064901",
	year = "2021"
}

@article{Ryu:2021lnx,
    author = "Ryu, Sangwook and Jupic, Vahidin and Shen, Chun",
    title = "{Probing early-time longitudinal dynamics with the \ensuremath{\Lambda} hyperon's spin polarization in relativistic heavy-ion collisions}",
    eprint = "2106.08125",
    archivePrefix = "arXiv",
    primaryClass = "nucl-th",
    doi = "10.1103/PhysRevC.104.054908",
    journal = "Phys. Rev. C",
    volume = "104",
    number = "5",
    pages = "054908",
    year = "2021"
}

@article{Ambrus:2020oiw,
    author = "Ambrus, Victor E. and Chernodub, M. N.",
    title = "{Hyperon\textendash{}anti-hyperon polarization asymmetry in relativistic heavy-ion collisions as an interplay between chiral and helical vortical effects}",
    eprint = "2010.05831",
    archivePrefix = "arXiv",
    primaryClass = "hep-ph",
    doi = "10.1140/epjc/s10052-022-10002-y",
    journal = "Eur. Phys. J. C",
    volume = "82",
    number = "1",
    pages = "61",
    year = "2022"
}

@article{Becattini:2020xbh,
    author = "Becattini, F. and Buzzegoli, M. and Palermo, A. and Prokhorov, G.",
    title = "{Polarization as a signature of local parity violation in hot QCD matter}",
    eprint = "2009.13449",
    archivePrefix = "arXiv",
    primaryClass = "hep-ph",
    doi = "10.1016/j.physletb.2021.136706",
    journal = "Phys. Lett. B",
    volume = "822",
    pages = "136706",
    year = "2021",
    note = "[Erratum: Phys.Lett.B 826, 136909 (2022)]"
}

@Article{Ivanov:2020udj,
  author        = {Ivanov, Yu B.},
  journal       = {Phys. Rev. C},
  title         = {{Global $\Lambda$ polarization in moderately relativistic nuclear collisions}},
  year          = {2021},
  number        = {3},
  pages         = {L031903},
  volume        = {103},
  archiveprefix = {arXiv},
  doi           = {10.1103/PhysRevC.103.L031903},
  eprint        = {2012.07597},
  primaryclass  = {nucl-th},
}

@Article{Guo:2021udq,
  author        = {Guo, Yu and Liao, Jinfeng and Wang, Enke and Xing, Hongxi and Zhang, Hui},
  journal       = {Phys. Rev. C},
  title         = {{Hyperon polarization from the vortical fluid in low-energy nuclear collisions}},
  year          = {2021},
  number        = {4},
  pages         = {L041902},
  volume        = {104},
  archiveprefix = {arXiv},
  doi           = {10.1103/PhysRevC.104.L041902},
  eprint        = {2105.13481},
  primaryclass  = {nucl-th},
}

@Article{Deng:2020ygd,
  author        = {Deng, Xian-Gai and Huang, Xu-Guang and Ma, Yu-Gang and Zhang, Song},
  journal       = {Phys. Rev. C},
  title         = {{Vorticity in low-energy heavy-ion collisions}},
  year          = {2020},
  number        = {6},
  pages         = {064908},
  volume        = {101},
  archiveprefix = {arXiv},
  doi           = {10.1103/PhysRevC.101.064908},
  eprint        = {2001.01371},
  primaryclass  = {nucl-th},
}

@article{Deng:2021miw,
    author = "Deng, Xian-Gai and Huang, Xu-Guang and Ma, Yu-Gang",
    title = "{Lambda polarization in 108Ag+108Ag and 197Au+197Au collisions around a few GeV}",
    eprint = "2109.09956",
    archivePrefix = "arXiv",
    primaryClass = "nucl-th",
    doi = "10.1016/j.physletb.2022.137560",
    journal = "Phys. Lett. B",
    volume = "835",
    pages = "137560",
    year = "2022"
}

@article{STAR:2021beb,
    author = "Abdallah, M. S. and others",
    collaboration = "STAR",
    title = "{Global $\Lambda$-hyperon polarization in Au+Au collisions at $\sqrt {s_{NN}}$=3~GeV}",
    eprint = "2108.00044",
    archivePrefix = "arXiv",
    primaryClass = "nucl-ex",
    doi = "10.1103/PhysRevC.104.L061901",
    journal = "Phys. Rev. C",
    volume = "104",
    number = "6",
    pages = "L061901",
    year = "2021"
}

@Article{Lei:2021mvp,
  author        = {Lei, Anke and Wang, Dujuan and zhou, Dai-Mei and Sa, Ben-Hao and Csernai, Laszlo Pal},
  journal       = {Phys. Rev. C},
  title         = {{Vorticity and \ensuremath{\Lambda} polarization in the microscopic transport model PACIAE}},
  year          = {2021},
  number        = {5},
  pages         = {054903},
  volume        = {104},
  archiveprefix = {arXiv},
  doi           = {10.1103/PhysRevC.104.054903},
  eprint        = {2110.13485},
  primaryclass  = {nucl-th},
}

@article{Hidaka:2022dmn,
    author = "Hidaka, Yoshimasa and Pu, Shi and Wang, Qun and Yang, Di-Lun",
    title = "{Foundations and applications of quantum kinetic theory}",
    eprint = "2201.07644",
    archivePrefix = "arXiv",
    primaryClass = "hep-ph",
    reportNumber = "KEK-TH-2390, J-PARC-TH-0267, RIKEN-iTHEMS-Report-22",
    doi = "10.1016/j.ppnp.2022.103989",
    journal = "Prog. Part. Nucl. Phys.",
    volume = "127",
    pages = "103989",
    year = "2022"
}

@Article{Monnai:2021kgu,
  author        = {Monnai, Akihiko and Schenke, Bj\"orn and Shen, Chun},
  journal       = {Int. J. Mod. Phys. A},
  title         = {{QCD Equation of State at Finite Chemical Potentials for Relativistic Nuclear Collisions}},
  year          = {2021},
  number        = {07},
  pages         = {2130007},
  volume        = {36},
  archiveprefix = {arXiv},
  doi           = {10.1142/S0217751X21300076},
  eprint        = {2101.11591},
  primaryclass  = {nucl-th},
}

@Article{Weil:2016zrk,
  author        = {Weil, J. and others},
  journal       = {Phys. Rev. C},
  title         = {{Particle production and equilibrium properties within a new hadron transport approach for heavy-ion collisions}},
  year          = {2016},
  number        = {5},
  pages         = {054905},
  volume        = {94},
  archiveprefix = {arXiv},
  doi           = {10.1103/PhysRevC.94.054905},
  eprint        = {1606.06642},
  primaryclass  = {nucl-th},
}

@article{Fu:2022myl,
    author = "Fu, Baochi and Pang, Longgang and Song, Huichao and Yin, Yi",
    title = "{Signatures of the spin Hall effect in hot and dense QCD matter}",
    eprint = "2201.12970",
    archivePrefix = "arXiv",
    primaryClass = "hep-ph",
    month = "1",
    year = "2022"
}

@article{Fu:2022oup,
    author = "Fu, Baochi and Pang, Longgang and Song, Huichao and Yin, Yi",
    title = "{Baryonic spin Hall effects in Au+Au collisions at $\sqrt{s_{NN}} = 7.7-200$ GeV}",
    eprint = "2208.00430",
    archivePrefix = "arXiv",
    primaryClass = "hep-ph",
    doi = "10.5506/APhysPolBSupp.16.1-A40",
    journal = "Acta Phys. Polon. Supp.",
    volume = "16",
    pages = "1--A40",
    year = "2023"
}

@Article{Yi:2021unq,
  author        = {Yi, Cong and Pu, Shi and Gao, Jian-Hua and Yang, Di-Lun},
  journal       = {Phys. Rev. C},
  title         = {{Hydrodynamic helicity polarization in relativistic heavy ion collisions}},
  year          = {2022},
  number        = {4},
  pages         = {044911},
  volume        = {105},
  archiveprefix = {arXiv},
  doi           = {10.1103/PhysRevC.105.044911},
  eprint        = {2112.15531},
  primaryclass  = {hep-ph},
}

@Article{Liang:2004ph,
  author        = {Liang, Zuo-Tang and Wang, Xin-Nian},
  journal       = {Phys. Rev. Lett.},
  title         = {{Globally polarized quark-gluon plasma in non-central A+A collisions}},
  year          = {2005},
  note          = {[Erratum: Phys.Rev.Lett. 96, 039901 (2006)]},
  pages         = {102301},
  volume        = {94},
  archiveprefix = {arXiv},
  doi           = {10.1103/PhysRevLett.94.102301},
  eprint        = {nucl-th/0410079},
  reportnumber  = {LBNL-56383},
}

@Article{Gao:2007bc,
  author        = {Gao, Jian-Hua and Chen, Shou-Wan and Deng, Wei-tian and Liang, Zuo-Tang and Wang, Qun and Wang, Xin-Nian},
  journal       = {Phys. Rev. C},
  title         = {{Global quark polarization in non-central A+A collisions}},
  year          = {2008},
  pages         = {044902},
  volume        = {77},
  archiveprefix = {arXiv},
  doi           = {10.1103/PhysRevC.77.044902},
  eprint        = {0710.2943},
  primaryclass  = {nucl-th},
  reportnumber  = {LBNL-63515},
}

@Article{Li:2017slc,
  author        = {Li, Hui and Pang, Long-Gang and Wang, Qun and Xia, Xiao-Liang},
  journal       = {Phys. Rev. C},
  title         = {{Global $\Lambda$ polarization in heavy-ion collisions from a transport model}},
  year          = {2017},
  number        = {5},
  pages         = {054908},
  volume        = {96},
  archiveprefix = {arXiv},
  doi           = {10.1103/PhysRevC.96.054908},
  eprint        = {1704.01507},
  primaryclass  = {nucl-th},
}

@Article{Vitiuk:2019rfv,
  author        = {Vitiuk, O. and Bravina, L. V. and Zabrodin, E. E.},
  journal       = {Phys. Lett. B},
  title         = {{Is different $\Lambda$ and $\bar \Lambda$ polarization caused by different spatio-temporal freeze-out picture?}},
  year          = {2020},
  pages         = {135298},
  volume        = {803},
  archiveprefix = {arXiv},
  doi           = {10.1016/j.physletb.2020.135298},
  eprint        = {1910.06292},
  primaryclass  = {hep-ph},
}

@Article{Fang:2016vpj,
  author        = {Fang, Ren-hong and Pang, Long-gang and Wang, Qun and Wang, Xin-nian},
  journal       = {Phys. Rev. C},
  title         = {{Polarization of massive fermions in a vortical fluid}},
  year          = {2016},
  number        = {2},
  pages         = {024904},
  volume        = {94},
  archiveprefix = {arXiv},
  doi           = {10.1103/PhysRevC.94.024904},
  eprint        = {1604.04036},
  primaryclass  = {nucl-th},
  reportnumber  = {ICTS-USTC-16-05},
}

@Article{Fang:2022ttm,
  author        = {Fang, Shuo and Pu, Shi and Yang, Di-Lun},
  journal       = {Phys. Rev. D},
  title         = {{Quantum kinetic theory for dynamical spin polarization from QED-type interaction}},
  year          = {2022},
  number        = {1},
  pages         = {016002},
  volume        = {106},
  archiveprefix = {arXiv},
  doi           = {10.1103/PhysRevD.106.016002},
  eprint        = {2204.11519},
  primaryclass  = {hep-ph},
}

@Article{STAR:2023eck,
  author        = {Abdulhamid, Muhammad and others},
  journal       = {Phys. Rev. Lett.},
  title         = {{Hyperon Polarization along the Beam Direction Relative to the Second and Third Harmonic Event Planes in Isobar Collisions at $\sqrt{s_\mathrm{NN}}$=200\,\,GeV}},
  year          = {2023},
  number        = {20},
  pages         = {202301},
  volume        = {131},
  archiveprefix = {arXiv},
  collaboration = {STAR},
  doi           = {10.1103/PhysRevLett.131.202301},
  eprint        = {2303.09074},
  primaryclass  = {nucl-ex},
}

@article{Yi:2023tgg,
    author = "Yi, Cong and Wu, Xiang-Yu and Yang, Di-Lun and Gao, Jian-Hua and Pu, Shi and Qin, Guang-You",
    title = "{Probing vortical structures in heavy-ion collisions at RHIC-BES energies through helicity polarization}",
    eprint = "2304.08777",
    archivePrefix = "arXiv",
    primaryClass = "hep-ph",
    doi = "10.1103/PhysRevC.109.L011901",
    journal = "Phys. Rev. C",
    volume = "109",
    number = "1",
    pages = "L011901",
    year = "2024"
}

@Article{Fang:2023bbw,
  author        = {Fang, Shuo and Pu, Shi and Yang, Di-Lun},
  journal       = {Phys. Rev. D},
  title         = {{Spin polarization and spin alignment from quantum kinetic theory with self-energy corrections}},
  year          = {2024},
  number        = {3},
  pages         = {034034},
  volume        = {109},
  archiveprefix = {arXiv},
  doi           = {10.1103/PhysRevD.109.034034},
  eprint        = {2311.15197},
  primaryclass  = {hep-ph},
}

@Article{HotQCD:2014kol,
  author        = {Bazavov, A. and others},
  journal       = {Phys. Rev. D},
  title         = {{Equation of state in ( 2+1 )-flavor QCD}},
  year          = {2014},
  pages         = {094503},
  volume        = {90},
  archiveprefix = {arXiv},
  collaboration = {HotQCD},
  doi           = {10.1103/PhysRevD.90.094503},
  eprint        = {1407.6387},
  primaryclass  = {hep-lat},
  reportnumber  = {BNL-105928-2014-JA},
}

@Article{McNelis:2021acu,
  author        = {McNelis, M. and Heinz, U.},
  journal       = {Phys. Rev. C},
  title         = {{Modified equilibrium distributions for Cooper--Frye particlization}},
  year          = {2021},
  number        = {6},
  pages         = {064903},
  volume        = {103},
  archiveprefix = {arXiv},
  doi           = {10.1103/PhysRevC.103.064903},
  eprint        = {2103.03401},
  primaryclass  = {nucl-th},
}

@Article{Lin:2024zik,
  author        = {Lin, Shu and Wang, Ziyue},
  title         = {{Steady state, displacement current and spin polarization for massless fermion in a shear flow}},
  year          = {2024},
  month         = {6},
  archiveprefix = {arXiv},
  eprint        = {2406.10003},
  primaryclass  = {hep-ph},
}

@article{Palermo:2024tza,
    author = "Palermo, Andrea and Grossi, Eduardo and Karpenko, Iurii and Becattini, Francesco",
    title = "{$\Lambda $ polarization in very high energy heavy ion collisions as a probe of the quark\textendash{}gluon plasma formation and properties}",
    eprint = "2404.14295",
    archivePrefix = "arXiv",
    primaryClass = "nucl-th",
    doi = "10.1140/epjc/s10052-024-13229-z",
    journal = "Eur. Phys. J. C",
    volume = "84",
    number = "9",
    pages = "920",
    year = "2024"
}

@Article{Becattini:2024uha,
  author        = {Becattini, Francesco and Buzzegoli, Matteo and Niida, Takafumi and Pu, Shi and Tang, Ai-Hong and Wang, Qun},
  title         = {{Spin polarization in relativistic heavy-ion collisions}},
  year          = {2024},
  month         = {2},
  archiveprefix = {arXiv},
  eprint        = {2402.04540},
  primaryclass  = {nucl-th},
}

@article{HADES:2022enx,
    author = "Abou Yassine, R. and others",
    collaboration = "HADES",
    title = "{Measurement of global polarization of \ensuremath{\Lambda} hyperons in few-GeV heavy-ion collisions}",
    eprint = "2207.05160",
    archivePrefix = "arXiv",
    primaryClass = "nucl-ex",
    doi = "10.1016/j.physletb.2022.137506",
    journal = "Phys. Lett. B",
    volume = "835",
    pages = "137506",
    year = "2022"
}

@article{STAR:2007ccu,
    author = "Abelev, B. I. and others",
    collaboration = "STAR",
    title = "{Global polarization measurement in Au+Au collisions}",
    eprint = "0705.1691",
    archivePrefix = "arXiv",
    primaryClass = "nucl-ex",
    reportNumber = "STAR-05-11-2007",
    doi = "10.1103/PhysRevC.76.024915",
    journal = "Phys. Rev. C",
    volume = "76",
    pages = "024915",
    year = "2007",
    note = "[Erratum: Phys.Rev.C 95, 039906 (2017)]"
}

@article{HotQCD:2018pds,
    author = "Bazavov, A. and others",
    collaboration = "HotQCD",
    title = "{Chiral crossover in QCD at zero and non-zero chemical potentials}",
    eprint = "1812.08235",
    archivePrefix = "arXiv",
    primaryClass = "hep-lat",
    doi = "10.1016/j.physletb.2019.05.013",
    journal = "Phys. Lett. B",
    volume = "795",
    pages = "15--21",
    year = "2019"
}

@article{Fang:2024vds,
    author = "Fang, Shuo and Pu, Shi",
    title = "{Collisional corrections to spin polarization from quantum kinetic theory using Chapman-Enskog expansion}",
    eprint = "2408.09877",
    archivePrefix = "arXiv",
    primaryClass = "hep-ph",
    doi = "10.1103/PhysRevD.111.034015",
    journal = "Phys. Rev. D",
    volume = "111",
    number = "3",
    pages = "034015",
    year = "2025"
}

@article{ParticleDataGroup:2020ssz,
    author = "Zyla, P. A. and others",
    collaboration = "Particle Data Group",
    title = "{Review of Particle Physics}",
    doi = "10.1093/ptep/ptaa104",
    journal = "PTEP",
    volume = "2020",
    number = "8",
    pages = "083C01",
    year = "2020"
}

@article{Lin:2025dsm,
    author = "Lin, Qian-Ru and Huang, Yu-Jing and Pang, Long-Gang and Luo, Xiao-Feng and Wang, Xin-Nian",
    title = "{Effects of Initial Nucleon-Nucleon Correlations on Light Nuclei Production in Au+Au Collisions at $\sqrt{s_\mathrm{NN}} = 3\ $ GeV}",
    eprint = "2503.01128",
    archivePrefix = "arXiv",
    primaryClass = "hep-ph",
    month = "3",
    year = "2025"
}

@article{Gutbrod:1989gh,
    author = "Gutbrod, H. H. and Kampert, K. H. and Kolb, B. and Poskanzer, Arthur M. and Ritter, H. G. and Schicker, R. and Schmidt, H. R.",
    title = "{Squeezeout of Nuclear Matter as a Function of Projectile Energy and Mass}",
    reportNumber = "LBL-28300",
    doi = "10.1103/PhysRevC.42.640",
    journal = "Phys. Rev. C",
    volume = "42",
    pages = "640--651",
    year = "1990"
}

@article{Bass:1993ce,
    author = "Bass, S. A. and Hartnack, C. and Stoecker, Horst and Greiner, W.",
    title = "{Out-of-plane pion emission in relativistic heavy ion collisions: Spectroscopy of Delta resonance matter}",
    doi = "10.1103/PhysRevLett.71.1144",
    journal = "Phys. Rev. Lett.",
    volume = "71",
    pages = "1144--1147",
    year = "1993"
}

@article{Hartnack:1994ce,
    author = "Hartnack, C. and Aichelin, J. and Stoecker, Horst and Greiner, W.",
    title = "{Out of plane squeeze of clusters in relativistic heavy ion collisions}",
    doi = "10.1016/0370-2693(94)90237-2",
    journal = "Phys. Lett. B",
    volume = "336",
    pages = "131--135",
    year = "1994"
}

@article{Sun:2025oib,
    author = "Sun, Kai-Jia and Liu, Dai-Neng and Zheng, Yun-Peng and Chen, Jin-Hui and Ko, Che Ming and Ma, Yu-Gang",
    title = "{Deciphering Hypertriton and Antihypertriton Spins from Their Global Polarizations in Heavy-Ion Collisions}",
    doi = "10.1103/PhysRevLett.134.022301",
    journal = "Phys. Rev. Lett.",
    volume = "134",
    number = "2",
    pages = "022301",
    year = "2025"
}

@article{Liu:2025kpp,
    author = "Liu, Dai-Neng and Zheng, Yun-Peng and Zhou, Wen-Hao and Chen, Jin-Hui and Ko, Che Ming and Ma, Yu-Gang and Sun, Kai-Jia and Zhang, Song",
    title = "{From Hyperons to Hypernuclei: A New Route to Unravel Proton Spin Polarization}",
    eprint = "2508.12193",
    archivePrefix = "arXiv",
    primaryClass = "nucl-th",
    month = "8",
    year = "2025"
}

@article{Zheng:2025ngn,
    author = "Zheng, Yun-Peng and Liu, Dai-Neng and Chen, Lie-Wen and Chen, Jin-Hui and Ko, Che Ming and Ma, Yu-Gang and Sun, Kai-Jia and Xu, Jun and Zhou, Bo",
    title = "{Global Spin Alignment of (Anti-)$^4$Li in Non-Central Heavy-Ion Collisions}",
    eprint = "2509.15286",
    archivePrefix = "arXiv",
    primaryClass = "nucl-th",
    month = "9",
    year = "2025"
}

@article{Xu:2026hxz,
    author = "Xu, Jun",
    title = "{Is nucleon spin thermalized in intermediate-energy heavy-ion collisions?}",
    eprint = "2602.23793",
    archivePrefix = "arXiv",
    primaryClass = "nucl-th",
    month = "2",
    year = "2026"
}

\end{document}